\documentclass[aps,prl,preprint,superscriptaddress,floatfix,showpacs]{revtex4-1}

\usepackage{graphicx}

\begin{document}


\title{Dynamics of colloidal particles in ice}


\author{Melissa Spannuth}
\email[]{melissa.spannuth@gmail.com}
\affiliation{Department of Geology and Geophysics, Yale University, New Haven, CT 06520, USA}
\author{S .G. J. Mochrie}
\affiliation{Department of Physics, Yale University, New Haven, CT 06520, USA}
\author{S. S. L. Peppin}
\affiliation{OCCAM, Mathematical Institute, University of Oxford, Oxford OX1 3LB, UK}
\author{J. S. Wettlaufer}
\affiliation{Department of Geology and Geophysics, Yale University, New Haven, CT 06520, USA}
\affiliation{Department of Physics, Yale University, New Haven, CT 06520, USA}
\affiliation{Program in Applied Mathematics, Yale University, New Haven, CT 06520, USA}


\date{\today}

\begin{abstract}

We use X-ray Photon Correlation Spectroscopy (XPCS) to probe the dynamics of colloidal particles in polycrystalline ice. During freezing, the dendritic ice morphology and rejection of particles from the ice created regions of high-particle-density, where some of the colloids were forced into contact and formed disordered aggregates. We find that the particles in these high density regions underwent ballistic motion coupled with both stretched and compressed exponential decays of the intensity autocorrelation function, and that the particles' characteristic velocity increased with temperature. We explain this behavior in terms of ice grain boundary migration. 

\end{abstract}


\maketitle


Solidification of the solvent phase of a colloidal suspension occurs in a wide variety of natural and technological settings. Owing to the rapidly expanding domain of materials applications, and relatively simple and inexpensive processing methods, a variety of solidification processes are receiving intense theoretical and experimental study \cite{deville2008}. Under a wide range of conditions, as the solvent freezes the interface between the liquid and solid solvent ramifies, guiding the particles into a variety of macroscopic morphologies. At the lowest freezing velocities all of the particles are rejected and pushed ahead of a macroscopically planar solid-liquid interface, while at higher velocities a dendritic solid-liquid interface aligns the particles into microporous structures, or solid lenses segregate the particles into periodic layers \cite{peppin2007, peppin2008}. At the highest growth rates the solid engulfs the particles individually. These various regimes have many applications including purifying water \cite{gay2002}, creating tissue scaffolds \cite{fu2008}, understanding ground freezing \cite{dash2006}, and building composite materials \cite{wilde2000}. The progression through the various regimes and their properties are fundamentally important, but involve a hierarchy of poorly understood cooperative phenomena. One of the factors limiting theoretical understanding is the dearth of experimental studies exploring the physical processes at work on the particle-scale. Here, we provide the first investigation of particle dynamics within frozen colloidal suspensions.

We used X-ray Photon Correlation Spectroscopy (XPCS) to examine the dynamic behavior of spherical silica colloids in ice. The scattering reveals ballistic dynamics combined with a non-exponential decay of the intensity autocorrelation function (ACF). This combination with a compressed exponential decay is commonly observed in light scattering from soft materials. Uniquely, in our experiments the decay is slower than exponential at small scattering vectors and faster than exponential at large scattering vectors. While ballistic dynamics combined with a stretched exponential have been reported previously in one other system, ours is evidently the first observation of a transition from stretched to compressed exponential behavior with increasing scattering vector. As such, it offers insight into the source of ballistic particle dynamics and non-exponential decay of the intensity ACF in non-equilibrium colloidal materials.


Samples of colloidal silica spheres (Polysciences, Inc.) dispersed in deionized water were solidified, and the composite material interrogated via X-ray scattering. The particle radius $R = 32$ nm, polydispersity $z = 18$\%, and initial unfrozen particle volume fraction $\phi \approx 2$\% were determined from small angle X-ray scattering (SAXS) \cite{spannuthsspreprint}. The samples were contained in an approximately $400 \, \mu$m thick, temperature-controlled sample chamber that produced a cylindrical isothermal region $2$ mm in diameter \cite{spannuthsspreprint}. Observations were made at several locations within the isothermal region of the samples.



The solvent was frozen by lowering the temperature at about $1^{\circ}$C/s. Freezing usually occurred near $T = -25^{\circ}$C. Therefore, the water in all samples was highly supercooled when ice growth began, resulting in an unstable solidification front and a cellular or dendritic ice growth morphology \cite{shibkov2003,peppin2007,deville2009}. Indeed, when we observed samples freezing under similar conditions with video microscopy the ice growth was dendritic \cite{spannuthsspreprint}. As expected \cite{deville2009}, the colloidal particles were rejected into the regions between the relatively pure ice dendrites resulting in mm-scale linear regions of high particle density tens of $\mu$m wide separated by regions of low particle density of approximately the same width \cite{spannuthsspreprint}.


After freezing, we increased $T$ in steps as small as $0.05^{\circ}$C from $T = -2^{\circ}$C to $T = 0^{\circ}$C, and obtained sets of images of the scattered X-ray intensity $I\left(\mathbf{q}, t\right)$ at various scattering vectors $\mathbf{q}$ and times $t$. We performed the X-ray scattering at sector 8-ID-I of the Advanced Photon Source \cite{lumma2000}. The beam size was $20 \, \mu$m by $20 \, \mu$m, ensuring that the illumnated volume of the sample contained many millions of particles. We used the multispeckle XPCS method to obtain the intensity autocorrelation function $g_2\left(q,\tau\right) = \langle I\left(q,t\right) \, I\left(q,t+\tau\right)\rangle / \langle I\left(q\right)\rangle^2$ from the azimuthally-averaged intensity $I\left(q,t\right)$. Here $\langle \cdots \rangle$ represents an ensemble average over the detector pixels and $\tau$ is the delay time between the two frames for which the correlation is calculated. The intensity ACF contains information about the particle dynamics through the shape and rate of its decay \cite{berneandpacora}. The length-scale of the dynamics probed is related to $q$; by using X-rays, we access length-scales comparable to the size of the individual particles. 

\begin{figure}[tp]
\begin{center}
\includegraphics[]{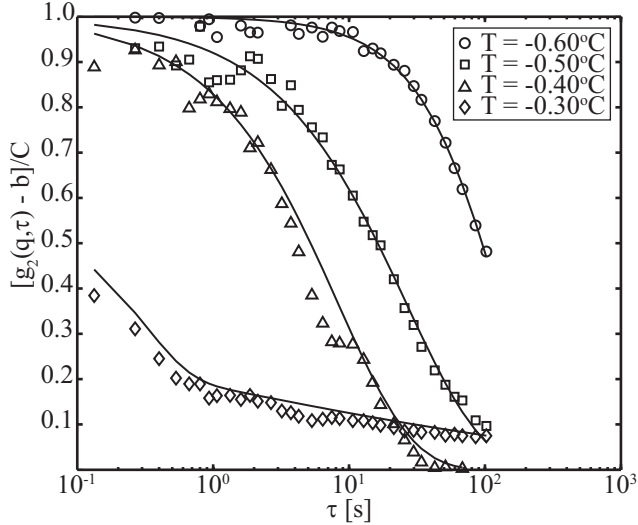} 
\caption[Intensity autocorrelation functions from XPCS]{Adjusted intensity ACFs $[g_2\left(q,\tau\right) - b ]/C$ versus delay time $\tau$ from sample $2$ (of four) at $T = -0.60^{\circ}$C (circles), $T = -0.50^{\circ}$C (squares), $T = -0.40^{\circ}$C (triangles), and $T = -0.30^{\circ}$C (diamonds) with the corresponding stretched or compressed exponential fits (solid curves).}
\label{XPCSg2s}
\end{center}
\end{figure}

Figure \ref{XPCSg2s} shows several ACFs at $q \approx 0.1$ nm$^{-1}$ (near the peak in the static scattered intensity) for a sample at $T = -0.60^{\circ}$C, $T = -0.50^{\circ}$C, $T = -0.40^{\circ}$C, and $T = -0.30^{\circ}$C. These data represent the typical forms of $g_2\left(q,\tau\right)$ observed: either a single decay, or two decays in which the second is extremely stretched and of small amplitude. For data with two decays, we are only concerned with the decay at small $\tau$. As the temperature increases, the decay rate also increases, so that the decay time becomes shorter.

In order to quantify the changes in decay time and shape, we fit each $g_2\left(q,\tau\right)$ to a Kohlrausch-Williams-Watts (KWW) expression \cite{williams1970} or a combination of two KWW expressions. Single decays were fit with $g_2\left(q,\tau\right) = b + C \exp \left[- 2 \left( \Gamma \tau\right)^{\alpha}\right]$, where $b$ is the baseline, $C$ is the contrast, $\Gamma$ is the decay rate, and $\alpha$ is the stretching ($< 1$) or compressing ($> 1$) exponent. Similarly, we fit double decays with $g_2\left(q,\tau\right) = 1 + C \left\{ \left( 1 - \beta \right) \exp \left[-\left(\Gamma_F \tau\right)^{\alpha_F}\right] + \beta \exp \left[-\left(\Gamma_S \tau\right)^{\alpha_S}\right] \right\}^2$, where $C$, $\Gamma$ and $\alpha$ are as above.  The ``partition coefficient'' $\beta$ describes the relative strength of the two exponential decays, and the subscripts denote the first ($F$) or second ($S$) decay. As illustrated by the solid curves in Fig. \ref{XPCSg2s}, most data were fit well with one of these functions, but some were too noisy or had features that did not fit into this analysis framework. We ascribe this variability to the spatial heterogeneity produced by the inherently stochastic nature of the ice nucleation process, the unstable ice growth morphology, and the process of ice crystal coarsening in the polycrystalline ice.

\begin{figure}[tp]
\begin{center}
\includegraphics[]{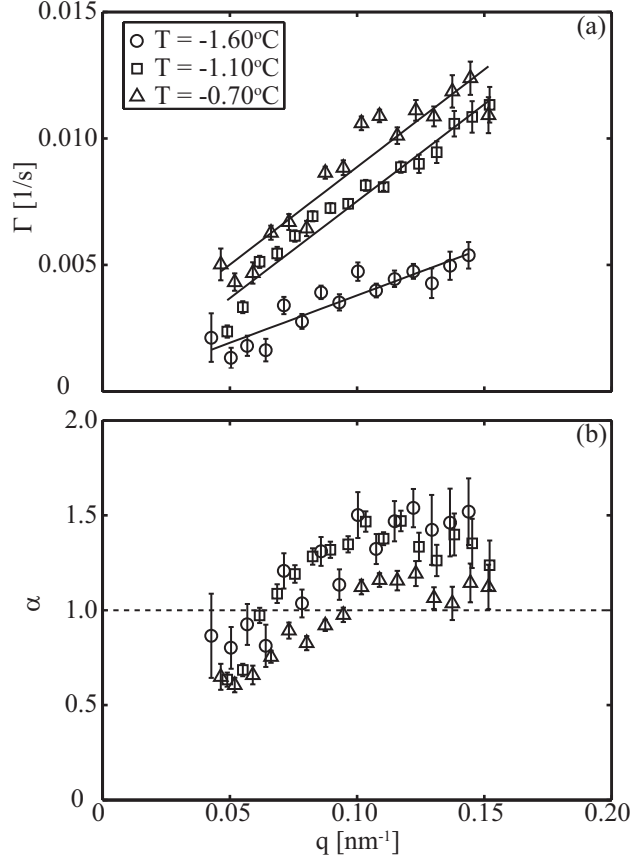} 
\caption[Gamma/alpha vs q]{In (a), examples of decay rate $\Gamma$ are plotted versus scattering vector $q$ from sample $2$ at $T = -1.60^{\circ}$C (circles), $T = -1.10^{\circ}$C (squares), and $T = -0.70^{\circ}$C (triangles) with their respective linear fits (solid lines). In (b), the corresponding exponents $\alpha$ are plotted versus $q$.}
\label{Gavsq}
\end{center}
\end{figure}

\begin{figure}[tp]
\begin{center}
\includegraphics[]{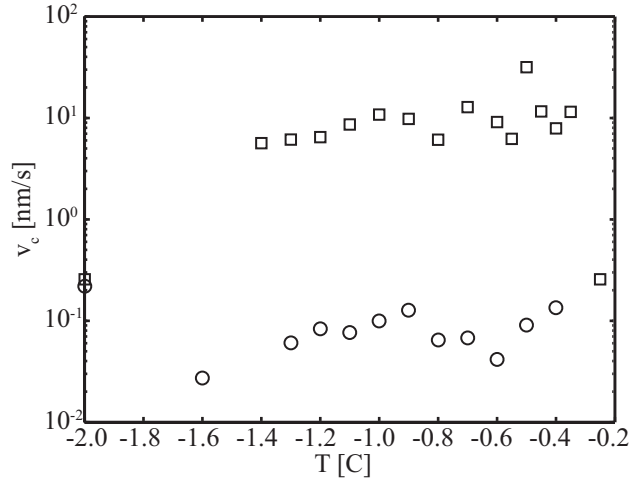} 
\caption[Characteristic velocity vs T]{The characteristic velocity $v_c$ versus temperature $T$ for sample $2$ (circles) and sample $3$ (squares).}
\label{charvelvsT}
\end{center}
\end{figure}

Both $\Gamma$ and $\alpha$ resulting from these fits vary with $q$. Figure \ref{Gavsq}a shows examples of $\Gamma$ versus $q$ for a sample at $T = -1.60^{\circ}$C, $T = -1.10^{\circ}$C, and $T = -0.70^{\circ}$C, all of which increase linearly with $q$. Figure \ref{Gavsq}b shows examples of $\alpha$ versus $q$ from the same data. In all cases, $\alpha$ is between $0.5$ and $1$ at low $q$ and increases to between $1$ and about $1.5$ at higher $q$, leveling off for $q \gtrsim 0.12$ nm$^{- 1}$. For comparison, the measured structure factor from SAXS has a peak at $q = 0.123$ nm$^{-1}$ and the particle size corresponds to $q = \pi / 32$ nm $= 0.0981$ nm$^{-1}$ \cite{spannuthsspreprint}. This behavior indicates ballistic particle dynamics and can be described with a distribution of particle velocities similar to a L{\'{e}}vy stable distribution, which has a power law tail \cite{cipelletti2003}. The characteristic velocity $v_c$ of the particles is the slope of a linear fit to $\Gamma$ versus $q$ (solid lines in Fig. \ref{Gavsq}a). We have fit $\Gamma$ from the data analyzed in this framework to find that $v_c$ generally increases with increasing temperature as shown in Fig. \ref{charvelvsT}. The orders of magnitude difference between the two curves can be ascribed to the variability introduced by the freezing process.

While ballistic dynamics combined with a compressed exponential decay of $g_2\left(q,\tau\right)$ has become a common observation \cite{cipelletti2003,bandy2004,duri2006,caronna2008}, the combination of ballistic dynamics with a stretched exponential decay is rare \cite{herzig2009} and a {\em transition} from {\em stretched} to {\em compressed} exponential decay with increasing scattering vector has not been reported previously. Either stretched or compressed decays could result from a continuous time random walk in which the size of the steps performed by the scatterers is distributed according to the L{\'{e}}vy alpha-stable distribution. This explanation is not specific to a particular material or system, but ascribing a physical origin to the neccessary step size distribution can be difficult. For colloidal gels, considering the relaxation of stress dipoles induced during disturbances to the material (e.g. shear or loading into a container) produces the values of $\alpha$ obtained experimentally \cite{bouchaud2002,duri2006}. However, few other systems exhibiting this type of behavior have such a complete description.

Here we suggest that the high-particle-density regions, which dominate the scattering, contained colloidal aggregates that formed when the solutions froze and that grain boundary motion arising from coarsening of the polycrystalline ice exerts forces on these aggregates resulting in the observed dynamic scattering. From SAXS performed simultaneously with XPCS, we know that the particles in the high density regions were close-packed and touching \cite{spannuthsspreprint}. The forces associated with particle rejection from the ice front during freezing are sufficient to overcome the electrostatic repulsion between the particles and bring them into contact \cite{rempel1999}. Once in contact, the van der Waal's attraction between the particles caused them to bind to each other forming an amorphous colloidal solid. The attractive energy for these particles is estimated to be $9 \times 10^{-20}$ J, or about $24 k_B T$ at $T = 273$ K, so thermal energy alone would be insufficient to completely disperse the aggregates on experimental time-scales. Indeed, using video microscopy we observed high-particle-density aggregates up to hundreds $\mu$m in size sedimenting through the melted samples. SAXS measurements indicated that about half of the particles initially in the solution sedimented \cite{spannuthsspreprint}. Thus, the grain boundaries between ice crystals likely contained a mixture of aggregates and single particles.

In order for the polycrystalline ice to coarsen, the aggregates and particles in the grain boundaries must deform under stresses transmitted from the ice crystals. We estimate the strength of these forces from the pressure $\Delta P$ driving the ice crystal coarsening: $\Delta P = \gamma_{gb} \kappa$, where $\gamma_{gb}$ is the surface free energy of the grain boundary and $\kappa$ is the curvature of the boundary \cite{cahn1962}. Taking $\gamma_{gb} = 0.029$ J$/$m$^2$, which is the value for an ice-water interface \footnote{Interstices between the particles should be water-filled at the $T$ studied due to curvature-induced depression of the melting temperature.}, and $\kappa = 1/R_c = 10^4$ m$^{-1}$, where $R_c = 100 \, \mu$m is an estimate of the ice grain boundary radius of curvature \footnote{Radius of curvature was estimated from the grain size in direct images.}, gives $\Delta P = 290$ N$/$m$^2$, which is distributed across the particles adjacent to the ice. Assuming that the pressure is distributed across about half the surface of a $10 \, \mu$m radius aggregate, the total force on the aggregate is about $3.6 \times 10^{-7}$ N. At $\phi = 0.6$, this amounts to a force per particle on the surface $F_p$ of about $2 \times 10^{-12}$ N. In order for the aggregate to deform, we assume that a particle must move a distance equal to one particle diameter ($64$ nm). The work performed by $F_p$ in moving a particle this distance is about $1 \times 10^{-19}$ J, slightly greater than the van der Waal's energy binding the particles together ($9 \times 10^{-20}$ J). Therefore, grain boundary motion induced by coarsening of the polycrystalline ice can deform the colloidal aggregates.

Although other stress sources are possible, grain boundary motion from coarsening produces characteristic velocities of the correct order of magnitude. The grain boundary velocity is given by $v_{gb} = M \gamma_{gb} \kappa$, where $M$ is the grain boundary mobility. The product $M \gamma_{gb}$ for ice at $T = -5^{\circ}$C ranges from about $10^{-11}$ to $10^{-14}$ m$^2/$s \cite{nasello2005}. Thus, for the value of $\kappa$ given above, $v_{gb}$ could be between $0.1$ and $100$ nm$/$s encompassing the observed range of $v_c$. Variations in grain size and mobility among the samples can account for the factor of $100$ difference between the characteristic velocities shown in Fig. \ref{charvelvsT}. Furthermore, grain boundary migration can account for the increase in the characteristic velocity with temperature through the Arrhenius behavior of the grain boundary mobility \cite{nasello2005}.


Determining how the stresses produced by grain boundary migration would affect the motion of particles in the grain boundaries is challenging; we discuss two complementary scenarios. It is unlikely that the stress resembles the dipolar stress sources proposed to explain compressed exponential decay and ballistic motion \cite{bouchaud2002}. Instead, the very high particle density may cause the colloids to behave more like a granular material (e.g., \cite{haw2004}) so that forces are distributed along force chains \cite{jaeger1996}. Such a direct transmission of the grain boundary motion would produce ballistic particle motion at about the same rate as the grain boundary was moving. However, the distribution of the force would cause particles to experience stress only sporadically. At larger length scales (low $q$), there would be a wider range of decay times (smaller $\alpha$) because the sporadic stress may be greater or less than that required to cause the necessary particle displacement for decorrelation of the intensity ACF. Whereas, at smaller length scales (large $q$) almost any stress would deform the particles sufficiently for decorrelation and thus there would be a narrower distribution of decay times (larger $\alpha$). However, the stress is still imposed sporadically leading to a non-negligible possibility of very long waiting times between imposition of sufficient stress, and hence a power law form.

We can connect this type of behavior with prior theoretical work on solidifying colloidal suspensions and self-filtration. In colloidal suspensions near random close packing, the osmotic pressure, and hence the generalized form of the Stokes-Einstein (mutual) diffusivity, diverge as $\phi \rightarrow \phi_{rcp}$ \cite{peppin2006, peppin2007, peppin2008}. Then, the character of diffusive relaxation of particle concentration gradients becomes extremely long ranged such that very near the divergent limit diffusion becomes effectively ``instantaneous.'' Thus, imposing a force on the close-packed colloids in the grain boundary leads to immediate and long-ranged particle motion, i.e. force chains. Such behavior of the diffusivity can be understood to drive ballistic motion.

Given the increasing study of high density and non-equilibrium systems, more examples of ballistic dynamics coupled with non-exponential decays are sure to emerge. Many of these will not fit into the developed framework and the manner of force transmission should be considered. For example, a similar type of stress distribution along force chains may explain the ballistic dynamics combined with a stretched exponential decay observed in particle-stabilized emulsions \cite{herzig2009}. Our experiments highlight the need for more general studies to quantify how different types of stresses manifest themselves in particle motion and scattering.




\begin{acknowledgments}
We thank S. Narayanan, A. Sandy and M. Sprung for assistance with the XPCS experiments, and X. Lu, J. Neufeld, E. Thomson and L. Wilen for useful discussions. MS acknowledges the NSF Graduate Research Fellowship for support. SGJM thanks the NSF for support via DMR-0906697. SSLP acknowledges support from KAUST Award KUK-C1-013-04. JSW acknowledges support from NSF Grant OPP0440841 and the US DoE Grant DE-FG02-05ER15741. Use of the Advanced Photon Source was supported by the US DoE under Contract DE-AC02-06CH11357.
\end{acknowledgments}



\begin{thebibliography}{27}%
\makeatletter
\providecommand \@ifxundefined [1]{%
 \@ifx{#1\undefined}
}%
\providecommand \@ifnum [1]{%
 \ifnum #1\expandafter \@firstoftwo
 \else \expandafter \@secondoftwo
 \fi
}%
\providecommand \@ifx [1]{%
 \ifx #1\expandafter \@firstoftwo
 \else \expandafter \@secondoftwo
 \fi
}%
\providecommand \natexlab [1]{#1}%
\providecommand \enquote  [1]{``#1''}%
\providecommand \bibnamefont  [1]{#1}%
\providecommand \bibfnamefont [1]{#1}%
\providecommand \citenamefont [1]{#1}%
\providecommand \href@noop [0]{\@secondoftwo}%
\providecommand \href [0]{\begingroup \@sanitize@url \@href}%
\providecommand \@href[1]{\@@startlink{#1}\@@href}%
\providecommand \@@href[1]{\endgroup#1\@@endlink}%
\providecommand \@sanitize@url [0]{\catcode `\\12\catcode `\$12\catcode
  `\&12\catcode `\#12\catcode `\^12\catcode `\_12\catcode `\%12\relax}%
\providecommand \@@startlink[1]{}%
\providecommand \@@endlink[0]{}%
\providecommand \url  [0]{\begingroup\@sanitize@url \@url }%
\providecommand \@url [1]{\endgroup\@href {#1}{\urlprefix }}%
\providecommand \urlprefix  [0]{URL }%
\providecommand \Eprint [0]{\href }%
\providecommand \doibase [0]{http://dx.doi.org/}%
\providecommand \selectlanguage [0]{\@gobble}%
\providecommand \bibinfo  [0]{\@secondoftwo}%
\providecommand \bibfield  [0]{\@secondoftwo}%
\providecommand \translation [1]{[#1]}%
\providecommand \BibitemOpen [0]{}%
\providecommand \bibitemStop [0]{}%
\providecommand \bibitemNoStop [0]{.\EOS\space}%
\providecommand \EOS [0]{\spacefactor3000\relax}%
\providecommand \BibitemShut  [1]{\csname bibitem#1\endcsname}%
\let\auto@bib@innerbib\@empty
\bibitem [{\citenamefont {Deville}(2008)}]{deville2008}%
  \BibitemOpen
  \bibfield  {author} {\bibinfo {author} {\bibfnamefont {S.}~\bibnamefont
  {Deville}},\ }\href@noop {} {\bibfield  {journal} {\bibinfo  {journal} {Adv.
  Eng. Mater}\ }\textbf {\bibinfo {volume} {10}},\ \bibinfo {pages} {155}
  (\bibinfo {year} {2008})}\BibitemShut {NoStop}%
\bibitem [{\citenamefont {Peppin}\ \emph {et~al.}(2007)\citenamefont {Peppin},
  \citenamefont {Worster},\ and\ \citenamefont {Wettlaufer}}]{peppin2007}%
  \BibitemOpen
  \bibfield  {author} {\bibinfo {author} {\bibfnamefont {S.~S.~L.}\
  \bibnamefont {Peppin}}, \bibinfo {author} {\bibfnamefont {M.~G.}\
  \bibnamefont {Worster}}, \ and\ \bibinfo {author} {\bibfnamefont {J.~S.}\
  \bibnamefont {Wettlaufer}},\ }\href@noop {} {\bibfield  {journal} {\bibinfo
  {journal} {Proc. R. Soc. A}\ }\textbf {\bibinfo {volume} {463}},\ \bibinfo
  {pages} {723} (\bibinfo {year} {2007})}\BibitemShut {NoStop}%
\bibitem [{\citenamefont {Peppin}\ \emph {et~al.}(2008)\citenamefont {Peppin},
  \citenamefont {Wettlaufer},\ and\ \citenamefont {Worster}}]{peppin2008}%
  \BibitemOpen
  \bibfield  {author} {\bibinfo {author} {\bibfnamefont {S.~S.~L.}\
  \bibnamefont {Peppin}}, \bibinfo {author} {\bibfnamefont {J.~S.}\
  \bibnamefont {Wettlaufer}}, \ and\ \bibinfo {author} {\bibfnamefont {M.~G.}\
  \bibnamefont {Worster}},\ }\href@noop {} {\bibfield  {journal} {\bibinfo
  {journal} {Phys. Rev. Lett.}\ }\textbf {\bibinfo {volume} {100}},\ \bibinfo
  {pages} {238301} (\bibinfo {year} {2008})}\BibitemShut {NoStop}%
\bibitem [{\citenamefont {Gay}\ and\ \citenamefont {Azouni}(2002)}]{gay2002}%
  \BibitemOpen
  \bibfield  {author} {\bibinfo {author} {\bibfnamefont {G.}~\bibnamefont
  {Gay}}\ and\ \bibinfo {author} {\bibfnamefont {M.~A.}\ \bibnamefont
  {Azouni}},\ }\href@noop {} {\bibfield  {journal} {\bibinfo  {journal} {Cryst.
  Growth Des.}\ }\textbf {\bibinfo {volume} {2}},\ \bibinfo {pages} {135}
  (\bibinfo {year} {2002})}\BibitemShut {NoStop}%
\bibitem [{\citenamefont {Fu}\ \emph {et~al.}(2008)\citenamefont {Fu},
  \citenamefont {Rahaman}, \citenamefont {Dogan},\ and\ \citenamefont
  {Bal}}]{fu2008}%
  \BibitemOpen
  \bibfield  {author} {\bibinfo {author} {\bibfnamefont {Q.}~\bibnamefont
  {Fu}}, \bibinfo {author} {\bibfnamefont {M.~N.}\ \bibnamefont {Rahaman}},
  \bibinfo {author} {\bibfnamefont {F.}~\bibnamefont {Dogan}}, \ and\ \bibinfo
  {author} {\bibfnamefont {B.~S.}\ \bibnamefont {Bal}},\ }\href@noop {}
  {\bibfield  {journal} {\bibinfo  {journal} {J. Biomed. Mater. Res. A}\
  }\textbf {\bibinfo {volume} {86B}},\ \bibinfo {pages} {125} (\bibinfo {year}
  {2008})}\BibitemShut {NoStop}%
\bibitem [{\citenamefont {Dash}\ \emph {et~al.}(2006)\citenamefont {Dash},
  \citenamefont {Rempel},\ and\ \citenamefont {Wettlaufer}}]{dash2006}%
  \BibitemOpen
  \bibfield  {author} {\bibinfo {author} {\bibfnamefont {J.~G.}\ \bibnamefont
  {Dash}}, \bibinfo {author} {\bibfnamefont {A.~W.}\ \bibnamefont {Rempel}}, \
  and\ \bibinfo {author} {\bibfnamefont {J.~S.}\ \bibnamefont {Wettlaufer}},\
  }\href@noop {} {\bibfield  {journal} {\bibinfo  {journal} {Rev. Mod. Phys.}\
  }\textbf {\bibinfo {volume} {78}},\ \bibinfo {pages} {695} (\bibinfo {year}
  {2006})}\BibitemShut {NoStop}%
\bibitem [{\citenamefont {Wilde}\ and\ \citenamefont
  {Perepezko}(2000)}]{wilde2000}%
  \BibitemOpen
  \bibfield  {author} {\bibinfo {author} {\bibfnamefont {G.}~\bibnamefont
  {Wilde}}\ and\ \bibinfo {author} {\bibfnamefont {J.~H.}\ \bibnamefont
  {Perepezko}},\ }\href@noop {} {\bibfield  {journal} {\bibinfo  {journal}
  {Mat. Sci. Eng. A - Struct.}\ }\textbf {\bibinfo {volume} {283}},\ \bibinfo
  {pages} {25} (\bibinfo {year} {2000})}\BibitemShut {NoStop}%
\bibitem [{\citenamefont {Spannuth}\ \emph {et~al.}(2010)\citenamefont
  {Spannuth}, \citenamefont {Mochrie}, \citenamefont {Peppin},\ and\
  \citenamefont {Wettlaufer}}]{spannuthsspreprint}%
  \BibitemOpen
  \bibfield  {author} {\bibinfo {author} {\bibfnamefont {M.~J.}\ \bibnamefont
  {Spannuth}}, \bibinfo {author} {\bibfnamefont {S.~G.~J.}\ \bibnamefont
  {Mochrie}}, \bibinfo {author} {\bibfnamefont {S.~S.~L.}\ \bibnamefont
  {Peppin}}, \ and\ \bibinfo {author} {\bibfnamefont {J.~S.}\ \bibnamefont
  {Wettlaufer}},\ }\href@noop {} {\bibfield  {journal} {\bibinfo  {journal}
  {arXiv:1011.1680v1 [cond-mat.soft]}\ } (\bibinfo {year} {2010})}\BibitemShut
  {NoStop}%
\bibitem [{\citenamefont {Shibkov}\ \emph {et~al.}(2003)\citenamefont
  {Shibkov}, \citenamefont {Golovin}, \citenamefont {Zheltov}, \citenamefont
  {Korolev},\ and\ \citenamefont {Leonev}}]{shibkov2003}%
  \BibitemOpen
  \bibfield  {author} {\bibinfo {author} {\bibfnamefont {A.~A.}\ \bibnamefont
  {Shibkov}}, \bibinfo {author} {\bibfnamefont {Y.~I.}\ \bibnamefont
  {Golovin}}, \bibinfo {author} {\bibfnamefont {M.~A.}\ \bibnamefont
  {Zheltov}}, \bibinfo {author} {\bibfnamefont {A.~A.}\ \bibnamefont
  {Korolev}}, \ and\ \bibinfo {author} {\bibfnamefont {A.~A.}\ \bibnamefont
  {Leonev}},\ }\href@noop {} {\bibfield  {journal} {\bibinfo  {journal}
  {Physica A}\ }\textbf {\bibinfo {volume} {319}},\ \bibinfo {pages} {65}
  (\bibinfo {year} {2003})}\BibitemShut {NoStop}%
\bibitem [{\citenamefont {Deville}\ \emph {et~al.}(2009)\citenamefont
  {Deville}, \citenamefont {Maire}, \citenamefont {Bernard-Granger},
  \citenamefont {Lasalle}, \citenamefont {Bogner}, \citenamefont {Gauthier},
  \citenamefont {Leloup},\ and\ \citenamefont {Guizard}}]{deville2009}%
  \BibitemOpen
  \bibfield  {author} {\bibinfo {author} {\bibfnamefont {S.}~\bibnamefont
  {Deville}}, \bibinfo {author} {\bibfnamefont {E.}~\bibnamefont {Maire}},
  \bibinfo {author} {\bibfnamefont {G.}~\bibnamefont {Bernard-Granger}},
  \bibinfo {author} {\bibfnamefont {A.}~\bibnamefont {Lasalle}}, \bibinfo
  {author} {\bibfnamefont {A.}~\bibnamefont {Bogner}}, \bibinfo {author}
  {\bibfnamefont {C.}~\bibnamefont {Gauthier}}, \bibinfo {author}
  {\bibfnamefont {J.}~\bibnamefont {Leloup}}, \ and\ \bibinfo {author}
  {\bibfnamefont {C.}~\bibnamefont {Guizard}},\ }\href@noop {} {\bibfield
  {journal} {\bibinfo  {journal} {Nat. Mater.}\ }\textbf {\bibinfo {volume}
  {8}},\ \bibinfo {pages} {966} (\bibinfo {year} {2009})}\BibitemShut {NoStop}%
\bibitem [{\citenamefont {Lumma}\ \emph {et~al.}(2000)\citenamefont {Lumma},
  \citenamefont {Lurio}, \citenamefont {Mochrie},\ and\ \citenamefont
  {Sutton}}]{lumma2000}%
  \BibitemOpen
  \bibfield  {author} {\bibinfo {author} {\bibfnamefont {D.}~\bibnamefont
  {Lumma}}, \bibinfo {author} {\bibfnamefont {L.~B.}\ \bibnamefont {Lurio}},
  \bibinfo {author} {\bibfnamefont {S.~G.~J.}\ \bibnamefont {Mochrie}}, \ and\
  \bibinfo {author} {\bibfnamefont {M.}~\bibnamefont {Sutton}},\ }\href@noop {}
  {\bibfield  {journal} {\bibinfo  {journal} {Rev. Sci. Instr.}\ }\textbf
  {\bibinfo {volume} {71}},\ \bibinfo {pages} {3274} (\bibinfo {year}
  {2000})}\BibitemShut {NoStop}%
\bibitem [{\citenamefont {Berne}\ and\ \citenamefont
  {Pecora}(1976)}]{berneandpacora}%
  \BibitemOpen
  \bibfield  {author} {\bibinfo {author} {\bibfnamefont {B.~J.}\ \bibnamefont
  {Berne}}\ and\ \bibinfo {author} {\bibfnamefont {R.}~\bibnamefont {Pecora}},\
  }\href@noop {} {\emph {\bibinfo {title} {Dynamic Light Scattering}}}\
  (\bibinfo  {publisher} {John Wiley and Sons},\ \bibinfo {year}
  {1976})\BibitemShut {NoStop}%
\bibitem [{\citenamefont {Williams}\ and\ \citenamefont
  {Watts}(1970)}]{williams1970}%
  \BibitemOpen
  \bibfield  {author} {\bibinfo {author} {\bibfnamefont {G.}~\bibnamefont
  {Williams}}\ and\ \bibinfo {author} {\bibfnamefont {D.}~\bibnamefont
  {Watts}},\ }\href@noop {} {\bibfield  {journal} {\bibinfo  {journal} {Trans.
  Faraday Soc.}\ }\textbf {\bibinfo {volume} {66}} (\bibinfo {year}
  {1970})}\BibitemShut {NoStop}%
\bibitem [{\citenamefont {Cipelletti}\ \emph {et~al.}(2003)\citenamefont
  {Cipelletti}, \citenamefont {Ramos}, \citenamefont {Manley}, \citenamefont
  {Pitard}, \citenamefont {Weitz}, \citenamefont {Pashkovski},\ and\
  \citenamefont {Johansson}}]{cipelletti2003}%
  \BibitemOpen
  \bibfield  {author} {\bibinfo {author} {\bibfnamefont {L.}~\bibnamefont
  {Cipelletti}}, \bibinfo {author} {\bibfnamefont {L.}~\bibnamefont {Ramos}},
  \bibinfo {author} {\bibfnamefont {S.}~\bibnamefont {Manley}}, \bibinfo
  {author} {\bibfnamefont {E.}~\bibnamefont {Pitard}}, \bibinfo {author}
  {\bibfnamefont {D.}~\bibnamefont {Weitz}}, \bibinfo {author} {\bibfnamefont
  {E.}~\bibnamefont {Pashkovski}}, \ and\ \bibinfo {author} {\bibfnamefont
  {M.}~\bibnamefont {Johansson}},\ }\href@noop {} {\bibfield  {journal}
  {\bibinfo  {journal} {Faraday Discuss.}\ }\textbf {\bibinfo {volume} {123}},\
  \bibinfo {pages} {237} (\bibinfo {year} {2003})}\BibitemShut {NoStop}%
\bibitem [{\citenamefont {Bandyopadhyay}\ \emph {et~al.}(2004)\citenamefont
  {Bandyopadhyay}, \citenamefont {Liang}, \citenamefont {Yardimci},
  \citenamefont {Sessoms}, \citenamefont {Borthwick}, \citenamefont {Mochrie},
  \citenamefont {Harden},\ and\ \citenamefont {Leheny}}]{bandy2004}%
  \BibitemOpen
  \bibfield  {author} {\bibinfo {author} {\bibfnamefont {R.}~\bibnamefont
  {Bandyopadhyay}}, \bibinfo {author} {\bibfnamefont {D.}~\bibnamefont
  {Liang}}, \bibinfo {author} {\bibfnamefont {H.}~\bibnamefont {Yardimci}},
  \bibinfo {author} {\bibfnamefont {D.~A.}\ \bibnamefont {Sessoms}}, \bibinfo
  {author} {\bibfnamefont {M.~A.}\ \bibnamefont {Borthwick}}, \bibinfo {author}
  {\bibfnamefont {S.~G.~J.}\ \bibnamefont {Mochrie}}, \bibinfo {author}
  {\bibfnamefont {J.~L.}\ \bibnamefont {Harden}}, \ and\ \bibinfo {author}
  {\bibfnamefont {R.~L.}\ \bibnamefont {Leheny}},\ }\href@noop {} {\bibfield
  {journal} {\bibinfo  {journal} {Phys. Rev. Lett.}\ }\textbf {\bibinfo
  {volume} {93}},\ \bibinfo {pages} {228302} (\bibinfo {year}
  {2004})}\BibitemShut {NoStop}%
\bibitem [{\citenamefont {Duri}\ and\ \citenamefont
  {Cipelletti}(2006)}]{duri2006}%
  \BibitemOpen
  \bibfield  {author} {\bibinfo {author} {\bibfnamefont {A.}~\bibnamefont
  {Duri}}\ and\ \bibinfo {author} {\bibfnamefont {L.}~\bibnamefont
  {Cipelletti}},\ }\href@noop {} {\bibfield  {journal} {\bibinfo  {journal}
  {Europhys. Lett.}\ }\textbf {\bibinfo {volume} {76}},\ \bibinfo {pages} {972}
  (\bibinfo {year} {2006})}\BibitemShut {NoStop}%
\bibitem [{\citenamefont {Caronna}\ \emph {et~al.}(2008)\citenamefont
  {Caronna}, \citenamefont {Chushkin}, \citenamefont {Madsen},\ and\
  \citenamefont {Cupane}}]{caronna2008}%
  \BibitemOpen
  \bibfield  {author} {\bibinfo {author} {\bibfnamefont {C.}~\bibnamefont
  {Caronna}}, \bibinfo {author} {\bibfnamefont {Y.}~\bibnamefont {Chushkin}},
  \bibinfo {author} {\bibfnamefont {A.}~\bibnamefont {Madsen}}, \ and\ \bibinfo
  {author} {\bibfnamefont {A.}~\bibnamefont {Cupane}},\ }\href@noop {}
  {\bibfield  {journal} {\bibinfo  {journal} {Phys. Rev. Lett.}\ }\textbf
  {\bibinfo {volume} {100}},\ \bibinfo {pages} {055702} (\bibinfo {year}
  {2008})}\BibitemShut {NoStop}%
\bibitem [{\citenamefont {Herzig}\ \emph {et~al.}(2009)\citenamefont {Herzig},
  \citenamefont {Robert}, \citenamefont {{van 't Zand}}, \citenamefont
  {Cipelletti}, \citenamefont {Pusey},\ and\ \citenamefont
  {Clegg}}]{herzig2009}%
  \BibitemOpen
  \bibfield  {author} {\bibinfo {author} {\bibfnamefont {E.~M.}\ \bibnamefont
  {Herzig}}, \bibinfo {author} {\bibfnamefont {A.}~\bibnamefont {Robert}},
  \bibinfo {author} {\bibfnamefont {D.~D.}\ \bibnamefont {{van 't Zand}}},
  \bibinfo {author} {\bibfnamefont {L.}~\bibnamefont {Cipelletti}}, \bibinfo
  {author} {\bibfnamefont {P.~N.}\ \bibnamefont {Pusey}}, \ and\ \bibinfo
  {author} {\bibfnamefont {P.~S.}\ \bibnamefont {Clegg}},\ }\href@noop {}
  {\bibfield  {journal} {\bibinfo  {journal} {Phys. Rev. E}\ }\textbf {\bibinfo
  {volume} {79}},\ \bibinfo {pages} {011405} (\bibinfo {year}
  {2009})}\BibitemShut {NoStop}%
\bibitem [{\citenamefont {Bouchaud}\ and\ \citenamefont
  {Pitard}(2002)}]{bouchaud2002}%
  \BibitemOpen
  \bibfield  {author} {\bibinfo {author} {\bibfnamefont {J.}~\bibnamefont
  {Bouchaud}}\ and\ \bibinfo {author} {\bibfnamefont {E.}~\bibnamefont
  {Pitard}},\ }\href@noop {} {\bibfield  {journal} {\bibinfo  {journal} {Eur.
  Phys. J. E}\ }\textbf {\bibinfo {volume} {9}},\ \bibinfo {pages} {287}
  (\bibinfo {year} {2002})}\BibitemShut {NoStop}%
\bibitem [{\citenamefont {Rempel}\ and\ \citenamefont
  {Worster}(1999)}]{rempel1999}%
  \BibitemOpen
  \bibfield  {author} {\bibinfo {author} {\bibfnamefont {A.~W.}\ \bibnamefont
  {Rempel}}\ and\ \bibinfo {author} {\bibfnamefont {M.~G.}\ \bibnamefont
  {Worster}},\ }\href@noop {} {\bibfield  {journal} {\bibinfo  {journal} {J.
  Cryst. Growth}\ }\textbf {\bibinfo {volume} {205}},\ \bibinfo {pages} {427}
  (\bibinfo {year} {1999})}\BibitemShut {NoStop}%
\bibitem [{\citenamefont {Cahn}(1962)}]{cahn1962}%
  \BibitemOpen
  \bibfield  {author} {\bibinfo {author} {\bibfnamefont {J.~W.}\ \bibnamefont
  {Cahn}},\ }\href@noop {} {\bibfield  {journal} {\bibinfo  {journal} {Acta
  Metall.}\ }\textbf {\bibinfo {volume} {10}},\ \bibinfo {pages} {789}
  (\bibinfo {year} {1962})}\BibitemShut {NoStop}%
\bibitem [{Note1()}]{Note1}%
  \BibitemOpen
  \bibinfo {note} {Interstices between the particles should be water-filled at
  the $T$ studied due to curvature-induced depression of the melting
  temperature.}\BibitemShut {Stop}%
\bibitem [{Note2()}]{Note2}%
  \BibitemOpen
  \bibinfo {note} {Radius of curvature was estimated from the grain size in
  direct images.}\BibitemShut {Stop}%
\bibitem [{\citenamefont {Nasello}\ \emph {et~al.}(2005)\citenamefont
  {Nasello}, \citenamefont {{Di Prinzio}},\ and\ \citenamefont
  {Gusm\'{a}n}}]{nasello2005}%
  \BibitemOpen
  \bibfield  {author} {\bibinfo {author} {\bibfnamefont {O.~B.}\ \bibnamefont
  {Nasello}}, \bibinfo {author} {\bibfnamefont {C.~L.}\ \bibnamefont {{Di
  Prinzio}}}, \ and\ \bibinfo {author} {\bibfnamefont {P.~G.}\ \bibnamefont
  {Gusm\'{a}n}},\ }\href@noop {} {\bibfield  {journal} {\bibinfo  {journal}
  {Acta Mater.}\ }\textbf {\bibinfo {volume} {53}},\ \bibinfo {pages} {4863}
  (\bibinfo {year} {2005})}\BibitemShut {NoStop}%
\bibitem [{\citenamefont {Haw}(2004)}]{haw2004}%
  \BibitemOpen
  \bibfield  {author} {\bibinfo {author} {\bibfnamefont {M.~D.}\ \bibnamefont
  {Haw}},\ }\href@noop {} {\bibfield  {journal} {\bibinfo  {journal} {Phys.
  Rev. Lett.}\ }\textbf {\bibinfo {volume} {92}},\ \bibinfo {pages} {185506}
  (\bibinfo {year} {2004})}\BibitemShut {NoStop}%
\bibitem [{\citenamefont {Jaeger}\ \emph {et~al.}(1996)\citenamefont {Jaeger},
  \citenamefont {Nagel},\ and\ \citenamefont {Behringer}}]{jaeger1996}%
  \BibitemOpen
  \bibfield  {author} {\bibinfo {author} {\bibfnamefont {H.~M.}\ \bibnamefont
  {Jaeger}}, \bibinfo {author} {\bibfnamefont {S.~R.}\ \bibnamefont {Nagel}}, \
  and\ \bibinfo {author} {\bibfnamefont {R.~P.}\ \bibnamefont {Behringer}},\
  }\href@noop {} {\bibfield  {journal} {\bibinfo  {journal} {Rev. Mod. Phys.}\
  }\textbf {\bibinfo {volume} {68}},\ \bibinfo {pages} {1259} (\bibinfo {year}
  {1996})}\BibitemShut {NoStop}%
\bibitem [{\citenamefont {Peppin}\ \emph {et~al.}(2006)\citenamefont {Peppin},
  \citenamefont {Elliot},\ and\ \citenamefont {Worster}}]{peppin2006}%
  \BibitemOpen
  \bibfield  {author} {\bibinfo {author} {\bibfnamefont {S.~S.~L.}\
  \bibnamefont {Peppin}}, \bibinfo {author} {\bibfnamefont {J.~A.~W.}\
  \bibnamefont {Elliot}}, \ and\ \bibinfo {author} {\bibfnamefont {M.~G.}\
  \bibnamefont {Worster}},\ }\href@noop {} {\bibfield  {journal} {\bibinfo
  {journal} {J. Fluid Mech.}\ }\textbf {\bibinfo {volume} {554}},\ \bibinfo
  {pages} {147} (\bibinfo {year} {2006})}\BibitemShut {NoStop}%
\end{thebibliography}

\end{document}